\documentclass[10pt]{article}

\usepackage{amsmath}
\usepackage{amsfonts}
\usepackage{amssymb}

\usepackage{graphicx}
\usepackage{subfigure}

\begin{document}

\newcommand{\be}{\begin{equation}}\newcommand{\ee}{\end{equation}}
\newcommand{\bea}{\begin{eqnarray}} \newcommand{\eea}{\end{eqnarray}}
\newcommand{\ba}[1]{\begin{array}{#1}} \newcommand{\ea}{\end{array}}

\numberwithin{equation}{section}

\def\np{Nucl. Phys. {\bf B}}\def\pl{Phys. Lett. {\bf B}}
\def\mpl{Mod. Phys. {\bf A}}\def\ijmp{Int. J. Mod. Phys. {\bf A}}
\def\cmp{Comm. Math. Phys.}\def\prd{Phys. Rev. {\bf D}}

\def\a{\alpha}
\def\b{\beta}
\def\g{\gamma}
\def\nn{\nonumber }
\def\eps{\epsilon }

\rightline{ICCUB-09-425}

\rightline{December, 2009}

\bigskip

\begin{center}

{\Large\bf  Models of Holographic superconductivity}
\bigskip
\bigskip

{\it \large Francesco Aprile\footnotemark[1] and Jorge G. Russo\footnotemark[1]\footnotemark[2]}
\bigskip

{\it
1) Institute of Cosmos Sciences and Estructura i Constituents de la Materia\\
Facultat de F{\'\i}sica, Universitat de Barcelona\\
Av. Diagonal 647,  08028 Barcelona, Spain\\
\smallskip
2) Instituci\'o Catalana de Recerca i Estudis Avan\c cats (ICREA)\\
Pg. Lluis Companys, 23, 08010 Barcelona, Spain\\
}
\bigskip
\bigskip

\end{center}
\bigskip

\begin{abstract}
We construct  general models for holographic superconductivity parametrized by
three couplings which are functions of a real scalar field and show that under general assumptions they describe superconducting phase transitions.
While some features are universal and model independent, important aspects of the quantum critical behavior strongly depend on
the choice of couplings, such as the order of the phase transition and critical exponents of second-order phase transitions.
In particular, we study a one-parameter model where the phase transition changes from second to first order above some critical value of
the parameter and a model with tunable critical exponents.

\end{abstract}

\clearpage

\tableofcontents

\section{Introduction}

Recently, the AdS/CFT correspondence \cite{malda,gubser,witten} was applied to study strongly coupled systems which undergo a superconducting
phase transition below a critical temperature \cite{Gubser:2008px,Hartnoll:2008vx,Gubser:2008zu,Hartnoll:2008kx,Gubser:AbHiggs}.
{}On the field theory side, in these models superconductivity is characterized by the condensation
of a composite charged operator  for low temperatures $T < T_c$. In the  dual 
gravitational description, the superconducting phase transition is represented by a transition from the black hole in anti-de Sitter space, with Hawking temperature equal to $T$, to
a new solution with scalar ``hair", which is thermodynamically preferred below the critical temperature $T_c$.

The models studied here are a  generalization of the models introduced in \cite{Franco:2009yz,Franco:2009if}.
In  particular cases, they reduce to the 3+1 dimensional model studied in \cite{Gauntlett:2009dn},
or the 4+1 dimensional model of \cite{Gubser:2009qm}.
Certain properties of superconductors, like the London equation or infinite DC conductivity, follow from $U(1)$ spontaneous symmetry breaking and are therefore universal (see e.g. \cite{weinberg}).
Other features, like the number of vacua and the detailed dynamics of the phase transition (including critical exponents), may be model dependent,  
as will be illustrated by means of some examples. 
Therefore the family of models constructed here could be used as a set-up for model building.

Another motivation in our construction is to have a framework where one can continuously interpolate between
different models appeared in the literature. Finally, we also hope that a special choice of couplings could lead to simpler
(perhaps analytical) solutions, and one example (albeit irrelevant for superconductivity) of a model
where the equations of motions are reduced to first-order differential equations is given in appendix A.

This paper is organized as follows. In section 2.1 we introduce the family of models and describe some  basic conditions that the  couplings should satisfy
in order to have AdS/CFT duality at work and a consistent model of holographic superconductivity. In section 2.2 we write down the ansatz and the equations of motion. 
We also describe the boundary conditions that give rise to the hairy black hole solutions that lead to a $U(1)$ spontaneous symmetry breaking in the boundary field theory.
In section 3 we present a preliminary discussion on how some aspects of the phase transitions should be encoded in the couplings. This is summarized  by
four conjectures.
We then present the numerical results for the integration of the differential equations (including the back reaction)
in three different models. These novel models  are designed to test the conjectures made at the beginning of the section.
The models of sect 3.1 and 3.2 give rise to  features which already appeared in inequivalent models in \cite{Franco:2009yz,Franco:2009if}, like the passage between first 
and second order phase transitions, and tunable critical exponents.
The model of section 3.3 describes a physics similar to the  model of Hartnoll, Herzog and Horowitz \cite{Hartnoll:2008kx} (hereafter HHH model). 
Section 4 contains concluding remarks. Appendix A gives the equations in a slightly more general gauge and
the expression for the conserved charge which is present for every model in our family. In this appendix we also show that
for a particular choice of couplings there is an extra conserved charge which represents an integral of a combination of the equations of motion.
This permits to write the equations of motion as first-order differential equations.
Finally,  appendix B
contains a simple derivation of the AdS Reissner-Nordstr\" om (RNAdS) solution
by the method of the superpotential.

\section{Generalized superconducting model}
\def\p{\partial }
\def\s{\sigma }

\subsection{The model}

Consider the following $U(1)$ invariant $d+1$ dimensional Lagrangian
\be
\sqrt{-g}{\cal L}= \sqrt{-g}\left(F(\eta) \ R -{1\over 4} G(\eta)\ F^{\mu\nu} F_{\mu\nu} - V(\eta) -{1\over 2} H(\eta) (\partial \eta)^2 -{1\over 2} J(\eta) (\partial_\mu \theta -A_\mu )^2\right)\ .
\label{cero}
\ee
with $\mu ,\ \nu =0,1,...,d$. 
It represents the most general covariant Lagrangian with local $U(1)$ symmetry, containing a metric $g_{\mu\nu}$, a $U(1)$ gauge field $A_\mu$, a real scalar field $\eta$ and a St\" uckelberg field $\theta $, and
terms with no more than two derivatives.
The gauge transformations are the standard ones given by
\be
A_\mu \to A_\mu +\partial_\mu \Lambda\ ,\qquad \theta\to\theta+\Lambda\ ,
\ee
with $g_{\mu\nu}$ and $\eta$ being invariant.

{}For $d>1$, by a suitable Weyl rescaling $g_{\mu\nu}\to f(\eta) g_{\mu\nu} $ and a field redefinition $\eta\to K(\eta)$, we can remove two of the couplings and put the Lagrangian in the following form
\be
\sqrt{-g} {\cal L}= \sqrt{-g}\left( \ R -{1\over 4} G(\eta)\ F^{\mu\nu} F_{\mu\nu} +{d(d-1)\over L^2} U(\eta) -{1\over 2} (\partial \eta)^2 -{1\over 2} J(\eta) A_\mu A^\mu\right)\ ,
\label{uno}
\ee
where we have also fixed the gauge $\theta =0$ and relabeled the couplings.
The Lagrangian (\ref{uno}) is our starting point for the construction of interesting models for holographic superconductors.
Our aim is to understand  the effect that $G(\eta),\ U(\eta) $ and $J(\eta )$ have on physical quantities and on the dynamics of the phase transition.

The requirement that the model  exhibits holographic superconductivity implies a number of
constraints on the behavior of the couplings $G,\ U,\ J$.
In particular, the couplings must be such that  the theory contains the usual $AdS$ solution.
This requires that there is an extrema of $U(\eta )$ at which $U(\eta )$ has a finite, positive value. By a suitable shift in $\eta $, with no loss of generality this extrema
can be placed at $\eta =0$ and $U(0)$ can be set to 1 (any other value of $U(0)>0$ can be absorbed into a redefinition of the parameter $L$).
Hence, at small $\eta $, one must have the behavior $U(\eta ) = 1+ O(\eta^2) $. The potential could also be such that it admits other extrema.
An example is the model with quartic potential $U(\eta)$   investigated in \cite{Gubser:AbHiggs,Gubser2}.
On the other hand, $G(\eta)$ and $J(\eta)$ must be positive 
to ensure that the kinetic terms for $A_\mu $ and $\theta $ have the correct signs.

We also request the theory to contain 
charged  black hole solutions with AdS asymptotics and $\eta =0$
(here we will not discuss models with dilatonic black holes that may also exhibit interesting critical phenomena, see \cite{gubser-dilatonic} for  some examples). 
This implies that  $G(0)$ must be finite and non-zero and with no loss of generality 
we can take $G(0)=1$ (which fixes the canonical normalization for $A_\mu$). 
In addition, the theory must also contain black hole solutions with non-trivial scalar hair $\eta $ with 
appropriate behavior at infinity, which implies that it must goes to zero in such a  way that the dual field operator takes an expectation value
(this point will be expanded below).
In addition, $J(\eta )$ must vanish as $\eta \to 0 $ since we want $U(1)$ to be unbroken at $\eta =0$.
The condition that the theory contains charged black hole solutions with $\eta =0$ also implies that $\partial_\eta J$ and $\partial_\eta G$
must vanish as $\eta \to 0 $, in order to satisfy the $\eta $ equation of motion.
Thus, in this paper, the couplings will be required to have the following behavior at small $\eta $:
\bea
G(\eta ) &=& 1+ c\eta^2+ O(\eta^3) \ ,
\nn\\
\label{asym}
U(\eta ) &=& 1-\frac{m^2}{2d(d-1)}\ \eta^2+ O(\eta^3)\ , 
\\
J(\eta ) &=& q^2\eta^2 + O(\eta^3) \ .
\nn
\eea
{} An important subclass are models with ${\bf Z}_2$ symmetry  $\eta \to -\eta $, for which the required behavior is
\bea
G(\eta ) &=& 1+ c\eta^2+ O(\eta^4) \ ,
\nn\\
\label{asymeven}
U(\eta ) &=& 1-\frac{m^2}{2d(d-1)}\ \eta^2+ O(\eta^4)\ , 
\\
J(\eta ) &=& q^2\eta^2 + O(\eta^4) \ .
\nn
\eea
i.e. the expansion contains only even powers of $\eta $.
The mass parameter $m^2$ is required to satisfy the Breitenlohner-Freedman bound for stability, $m^2\geq -d^2/4$.
It should be noted that it is not necessary to have $q\neq 0$ in order to have holographic superconductivity. In particular, in \cite{Franco:2009yz} it
was shown that the choice $J= \eta^3 $  leads to superconductivity as well.

Different choices of $G\ ,\ U,\ , J$ reproduce the various  models that appeared in the literature.
In particular, the HHH  model \cite{Hartnoll:2008kx} corresponds to taking $d=3$ and the choice
\be
G(\eta )=1\ ,\qquad U(\eta ) = 1 + {1\over 6}\ \eta^2\ ,\qquad J(\eta ) = q^2 \eta^2\ .
\label{stm}
\ee
{} For a general model with the asymptotic (\ref{asym}) and $m^2=-2$, the onset of instabilities will occur at the same critical temperature as 
in \cite{Hartnoll:2008kx}, since in this case the linearized Lagrangian is the same.

Other $d=3$ models that appeared in the literature are obtained by the following identifications.
The model of Franco et al \cite{Franco:2009yz} is the choice
\be
G(\eta )=1\ ,\qquad U(\eta ) = 1+ {1\over 6}\ \eta^2\ ,\qquad J(\eta ) = {\rm arbitrary}\ .
\ee
The M-theory model of Gauntlett et al \cite{Gauntlett:2009dn} corresponds to taking
\be
G(\eta )=1\ ,\qquad U(\eta ) = {1\over 6} \cosh^2{\eta\over 2}\ (7-\cosh\eta ) \ ,\qquad J(\eta ) = {1\over L^2}\ \sinh^2\eta\ .
\ee
The string theory model of Gubser et al \cite{Gubser:2009qm} is obtained by considering  the same Lagrangian (\ref{uno}) with $d=4$ and setting
\be
G(\eta )=1\ ,\qquad U(\eta ) = {1\over 2} \cosh^2{\eta\over 2}\ (5-\cosh\eta ) \ ,\qquad J(\eta ) = {3\over L^2}\ \sinh^2\eta\ .
\ee

An even wider class of models can be obtained if one defines $\eta =e^\varphi $, with the fundamental degree of freedom  being $\varphi $, a real scalar field. 
In other words, if we assume that $\eta >0$. This permits a more general dependence on $\eta $, that includes non-integer powers.
In particular, the couplings would now be required to have the asymptotic behavior
\bea
G(\eta ) &=& 1+ O\big(\eta^{\alpha_1} \big) \ ,
\nn\\
\label{asym2}
U(\eta ) &=& 1-\frac{m^2}{2d(d-1)}\ \eta^2+ O\big( \eta^{\alpha_2}\big)\ , 
\\
J(\eta ) &=& q^2\eta^2 + O\big( \eta^{\alpha_3}\big) \ .
\nn
\eea
with $\alpha_1,\ \alpha_2,\ \alpha_3$ being any real numbers subject to  the condition $\alpha_1>1,\alpha_{2,3}>2$
(the condition $\alpha_1>1$ comes from the requirement that the RNAdS black hole with $\eta=0$ solves the $\eta $ equation).  
A particular example of this is the $d=3$ model investigated in \cite{Franco:2009if}, which in our notation reads
\be
G(\eta)=1\ ,\qquad U(\eta)= 1 + {1\over 6}\ \eta^2\ ,\qquad J(\eta ) =  \eta^2 +c_\alpha \eta^\alpha +c_4 \eta^4\ , \quad 3\leq \alpha\leq 4\ .
\label{tuna}
\ee
This model exhibits interesting features like tunable critical exponents, i.e. which change by varying  the parameter $\alpha $.
\medskip

Thus the Lagrangian (\ref{uno}) provides a universal structure that embodies various models studied in the literature and permits a more general approach
to the phenomenology of superconductors.


%

\subsection{Holographic superconductivity}

We now set $d=3$ and consider the following ansatz:
\be
ds^2= -g(r) e^{-\chi (r)} dt^2 +{dr^2\over g(r)} + r^2(dx^2+dy^2)\ ,\qquad A=\phi(r)dt\ ,\qquad \eta=\eta(r)\ .
\label{arass}
\ee
The effective Lagrangian takes the form
\be\label{L1}
\sqrt{-g} {\cal L} = -2 e^{-{\chi\over 2}} (r g)'+ {r^2\over 2} G(\eta ) e^{\chi\over 2} {\phi'}^2+{6r^2\over L^2}  e^{-{\chi\over 2}} U(\eta ) -{r^2\over 2} e^{-{\chi\over 2}} g {\eta'}^2
+{r^2\over 2g} e^{\chi \over 2} J(\eta ) \phi^2,
\ee
and the equations of motion reduce to
\bea
&&\chi'+ {r\over 2} {\eta'}^2+ {r\over 2g^2} e^\chi J(\eta)\phi^2=0\ ,
\label{buno}\\
&& {1\over 4}\ {\eta'}^2+ {G(\eta)\over 4g}\ e^\chi {\phi'}^2+{g'\over rg} +{1\over r^2} -{3\over L^2 g}\ U(\eta) +{1\over 4g^2}\ e^\chi J(\eta) \phi^2=0\ ,
\label{bdos}\\
&& \phi''+\phi' \ \left( {2\over r} +{\chi'\over 2} +{\p_\eta G \eta'\over G} \right)- {J(\eta)\over g G(\eta)} \ \phi =0\ ,
\label{btres}\\
&& \eta'' +\eta'  \ \left( {2\over r} -{\chi'\over 2} +{g'\over g} \right)+ {1\over 2g}\ e^\chi \p_\eta G \ {\phi'}^2 + {6\over L^2g}\ \p_\eta U + {1\over 2g^2}\ e^\chi \p_\eta J\ {\phi}^2 =0\ .
\label{bcuatro}
\eea
In the appendix A we give the equations in a more general coordinate system and discuss the presence of Noether charges.

The basic strategy to solve these equations is explained in detail in \cite{Hartnoll:2008kx,Gubser:AbHiggs}. Here we summarize the main features.
Black holes with  regular event horizon are needed in order to put the field theory at finite temperature. 
We require that $g$ has a simple zero at the event horizon $r=r_+$, that is, $g(r_+)=0$ but $g'(r_+)\neq 0$ and finite. 
The Hawking temperature of the black hole is, 
\be
T_{\rm Hawk} =\frac{1}{4\pi} g'(r)e^{-\chi(r)/2}\Big|_{r=r_+} \ .
\label{hawkt}
\ee
On the other hand, $g'(r_+)$ is not an independent parameter in our gauge but it is determined by a combination of eqs. (\ref{buno}) and (\ref{bdos}),
\be
\left( r g e^{-{\chi\over 2}}\right)' = {3 r^2\over L^2}  e^{-{\chi\over 2}}U(\eta) -   { r^2G(\eta)\over 4}  \        e^{{\chi\over 2}}{\phi'}^2\ .
\ee
Substituting this value of $g'$ into the Hawking temperature (\ref{hawkt}) we find 
\be
T_{\rm Hawk} = {r_+\over 16\pi L^2}\left( e^{-{\chi_+\over 2}} 12 U(\eta_+) - L^2 e^{\chi_+\over 2}G(\eta_+) \ E_+^2 \right),
\label{HK}
\ee
where we have defined
\be
\eta(r_+)\equiv\eta_+,\qquad \phi'(r_+)\equiv E_+,\qquad \chi(r_+)\equiv\chi_+\ .
\ee
Solving the full set of equations involves numerical integration from the horizon to infinity. Therefore, in order to have a well-posed Cauchy problem, 
we have to specify the remaining initial data at $r=r_+$. By expanding   (\ref{bcuatro}) in the vicinity of the horizon one 
determines the value of $\eta'(r_+)=\eta'_+$ through the relation,
\be
\eta'_+ g'(r_+) + {1\over 2} \ e^{\chi_+}\p_\eta G(\eta_+)\ E_+^2+{6\over L^2}\p_\eta U(\eta_+) = 0\ .
\ee
In addition, the condition $\phi(r_+)=0$ is needed to ensure that the gauge one-form is well-defined at the horizon  (see \cite{Gubser:2008px} for a discussion).
The solution is therefore specified by the values $r_+$, $\eta_+$, $E_+$ and $\chi_+$. This set of parameters can be  further reduced 
by using  two scaling symmetries of the metric, the gauge field and the equations of motion. The scaling dimensions are summarized in the following table:
\be\label{table}
\begin{array}{c|c|c|c|c|c|c}
	\mathrm{Symm.}&e^{\chi}&\ t\ &\ \vec x &r&g&\ \phi\\
	\hline\hline
	\rule{0pt}{4mm}\mathrm{I}&2&\phantom{-}1\ &\phantom{-}0\ &0&0&-1\\
	\mathrm{II}&0&-1\ &-1\ &1&2&\phantom{-}1
\end{array}	
\ee
The  symmetry II can be used to set $r_+=1$. 
The symmetry I can be used to set $\chi_0 =0$, where $\chi_0$ is the asymptotic value of $\chi $.
Thus the initial value data is completely characterized in terms of two parameters,  $\eta_+$ and $E_+$.

Because of the  conditions (\ref{asym}), the asymptotic behavior of the solution
will be essentially the same as in the HHH model. At infinity, the solution approaches the AdS geometry with
\be
g=\frac{r^2}{L^2}+\ldots
\ee 
{} Using  (\ref{btres})-(\ref{bcuatro}) one can then find the asymptotic expressions for $\phi$ and $\eta$,
\be
\phi(r)=\mu-\frac{\rho}{r}+\ldots ,\qquad\qquad \eta(r)=\frac{\eta^{(1)}}{r^{\lambda_-}}+\frac{\eta^{(2)}}{r^{\lambda_+}}+\ldots ,
\ee
Here $\lambda_{\pm}$ are the roots of $\lambda(\lambda-3)=m^2$.
The gauge-gravity correspondence proceeds as usual (see e.g.
\cite{Klebanov:1999tb,Son:2002sd}). 
The scalar field is dual to an operator $O_{\Delta}$ of dimension $\Delta$ equal to the larger root $\lambda_+$, while $\mu$ and $\rho$ are interpreted as the chemical potential and the charge density of the dual field theory. In order to have spontaneous symmetry breaking in the field theory, we have to consider solutions with $\eta^{(1)}=0$. This condition is, in fact,  equivalent to demanding that $O_{\lambda_+}$ is not sourced.
{} For $m^2$ in the interval $-9/4<m^2<-5/4 $ it is also consistent to swap  the role of $O_{\lambda_+}$ and $O_{\lambda_-}$ as both modes with $\eta^{(1)}$ or $\eta^{(2)}$ are normalizable. 

The set of pairs $(\eta_+,E_+)$ solving the condition $\eta^{(1)}(\eta_+,E_+)=0$ define the condensed phase in which $\langle O_{\lambda_+}\rangle \neq 0$.
As a result of this condition, the hairy black hole solution is fully characterized by a single parameter, which can be taken to be the temperature.


The non-condensed phase is obviously defined by $\langle O_{\Delta}\rangle=0$, i.e. $\eta^{(1)}=\eta^{(2)}=0$. 
This condition implies $\eta=0$, therefore this phase is dual to the usual AdS Reissner-Nordstr\" om  solution
\be
g= r^2 - {1\over r}\left( r_+^3+{\rho^2\over 4 r_+}\right) +{\rho^2\over 4 r^2}\ ,\qquad \phi=\rho\left ({1\over r_+}-{1\over r}\right)\ ,\qquad \chi = 0\ ,
\label{RNg}
\ee
in units where $L=1$ (see appendix B for a derivation).

Superconductivity only exists if the condensed phase is thermodynamically preferred, namely if the difference of free energy,
\be
\Delta f(T)= f_{\mathrm{cond}}(T)-f_{\mathrm{uncond}}(T)\ ,
\ee
is negative for some range of temperatures. In the next section we will see  typical examples where, at sufficiently low temperatures,  $\Delta f$ is  negative and the system is in a superconductor state; then it increases as the temperature is increased  up to some critical value $T=T_c$ at which $\Delta f(T_c)=0$. For temperatures above $T_c$
the uncondensed solution has less free energy and becomes the thermodynamically favorable;  the system turns into a normal state. 
The behavior of $\Delta f$ at $T=T_c$ determines the type of phase transition, according to the standard  classification.
For our system the free energy can be expressed in terms of the energy density, chemical potential and density charge. 
The energy density of the field theory configuration is associated with the mass $\epsilon$ of the black hole solution,
 which can be read from the asymptotic expression for the 00 component of the metric,
\be
e^{-\chi}g={r^2}-\frac{\epsilon }{2r}+\ldots
\ee 
A similar calculation as in
\cite{Hartnoll:2008vx}  gives
\be
f=-\frac{\epsilon}{2}+\mu\rho\ .
\ee
for all $(G,U,J)$ models. We will study the temperature dependence of the free energy in different models by working at fixed charge density $\rho $.

\section{Numerical Results for three examples}

General models with couplings that have the asymptotic behavior (\ref{asym}) 
are naively expected to exhibit similar dynamics near the critical point, where $\eta $ is small and higher order terms in the expansion in powers of $\eta $ are
in principle irrelevant. 
In particular, this expectation is partly confirmed by results
obtained in \cite{Gubser:2008px,Hartnoll:2008vx,Gubser:AbHiggs}, which show that 
the details of the potential $U$ are irrelevant as far as the existence of a condensed phase is concerned.
The mass parameter in $U=1 - m^2 \eta^2/12+...$ affects the critical temperature but the qualitative features of the transition
are  essentially the same \cite{Horowitz:2008bn}.
Moreover, it was shown in \cite{Maeda:2009wv} that for the HHH model the critical exponents take the standard mean field values.

\smallskip 

Nonetheless, this picture does not apply for all models. First, 
as shown in \cite{Franco:2009yz} for the $J=\eta^3 $ model,
for certain models the transition can be first order and occur at a finite (non-small) value
of $\eta$.\footnote{
In this paper, back reaction on the geometry is neglected --an assumption that can be justified to some extent by taking a suitably limit involving large $c_3$ in a $J=c_3\eta^3 $ model.
Thus a natural question is whether these important changes in the dynamics of the phase transition 
might be an artifact of the no back-reaction approximation. 
We have checked that the study of the phase transition including the complete back-reaction reproduces essentially the same features observed in \cite{Franco:2009yz}, even for low values of $c_3$ . 
}
The first example discussed here will illustrate this fact. It consists of a model obtained by a choice of non-trivial, one-parameter
coupling $G( \eta)$ in the Maxwell kinetic term, while having the same couplings  $U(\eta )$ and $J(\eta )$ as in the HHH model.
The model exhibits a number of interesting features that we discuss.
Second, there are also models exhibiting second-order transitions where critical exponents are modified.
An example is the model (\ref{tuna}) studied in \cite{Franco:2009if}.
The critical exponent $\beta $ for the order parameter $\langle O_2 \rangle \sim (T_c-T)^\beta$ is given by the relation $\beta =(\alpha-2)^{-1}$.
Although $\eta $ is very small near the transition, the term $\eta^\alpha $ in (\ref{tuna}) is not sufficiently suppressed and it affects the critical behavior
near $T_c$. 

\medskip

By an intuition based on the few known examples, one may conjecture that 

\smallskip

\noindent 1) For ${\bf Z}_2$-symmetric models with the asymptotic behavior (\ref{asymeven}), with $q\neq 0$, there is a certain range of parameters where they undergo a second-order phase transition. They have
the standard  critical exponents predicted by mean field theory.

\smallskip

\noindent 2) In these models the critical temperature of a second-order phase transition is sensitive to the values of the parameters $c,\ m,\ q$ and insensitive to the coefficients of the following powers.

\smallskip

\noindent 3) For  models which are not ${\bf Z}_2$ symmetric, with the asymptotic behavior (\ref{asym}), and $q\neq 0$, the critical exponents of second order transitions
can be affected by the powers $O(\eta^3)$ in (\ref{asym}) but they are insensitive to the next terms starting with $O(\eta^4)$.

\smallskip

\noindent 4) Models with $\eta >0$, i.e. having a real scalar field $\varphi$ as a fundamental variable, defined by $\eta \equiv e^\varphi >0$, 
can have the asymptotic behavior (\ref{asym2}). In a regime of parameters where
the couplings  $G$, $U$ and $J$ are congruent to (\ref{stm}), these models have second-order phase transitions
with  critical exponents that can depend on the values of the parameters.

\smallskip

The examples given below, with three different choices of $G(\eta )$, partly confirm this picture (and complement the study of \cite{Franco:2009yz,Gauntlett:2009dn,Franco:2009if},
performed for $G=1$ and different $U(\eta ),\ J(\eta )$ couplings).

\subsection{Model with $G(\eta )=(1+b\eta^2)^{-1}$ }

Consider a model with the following couplings
\be
G=\frac{1}{1+b\eta^2} \ ,\qquad U=1+{\eta^2\over 6}\ ,\qquad  J=q^2\eta^2\ .
\ee
The special value $b=0$ corresponds to  the HHH model.
The theory contains a variety of black hole solutions.
The uncondensed $\eta=0$  phase is described as usual by the RNAdS solution (\ref{RNg}).
When the temperature is below some critical value $T_c$, one or more hairy black hole solutions appear, depending on the value of the parameter $b$.
{} For $b$ below some critical value $b_{\rm cr}\sim 3$, there is a single black hole with hair.
When $b$ is greater than $b_{\rm cr}$,  new hairy black holes appear.
Remarkably, the picture turns out to be qualitatively similar as in the model discussed in \cite{Franco:2009if}
with $G=1$, $U= 1 + {1\over 6}\ \eta^2 $,  and $J= \eta^2+ c_4 \eta^4$, with  $c_4$ playing the role of $b$. One difference is that
in the present case the critical temperature depends on $b$ (in consistency with conjecture 2), whereas for the model of \cite{Franco:2009if} with 
$J= \eta^2+ c_4 \eta^4$ the critical temperature remains fixed and independent of $c_4$ (which is also a prediction of conjecture 2).

\begin{figure}[tbh]
\centering
\subfigure[]{\includegraphics[width=7.5cm]{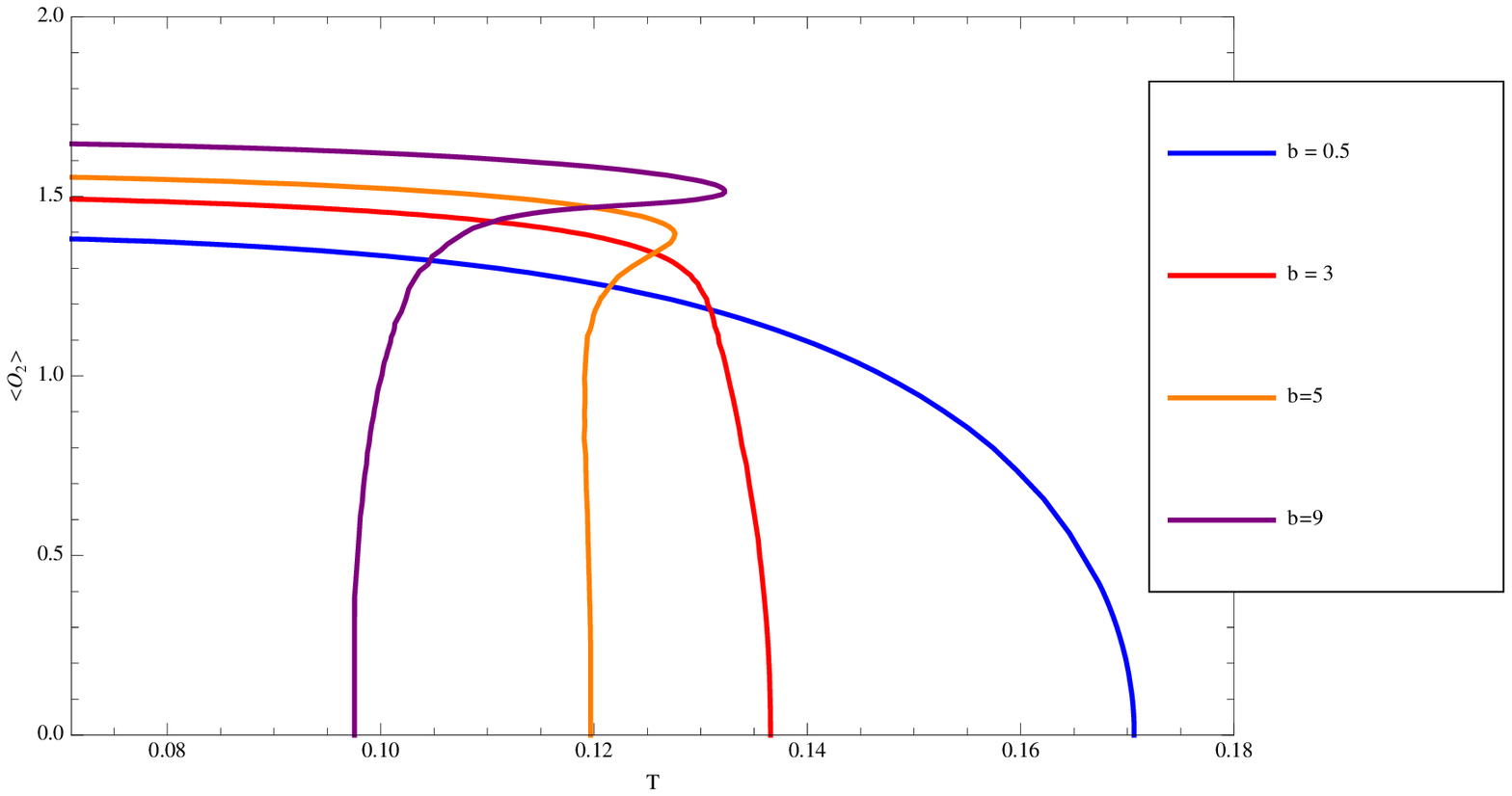}}
\subfigure[]{\includegraphics[width=7.5cm]{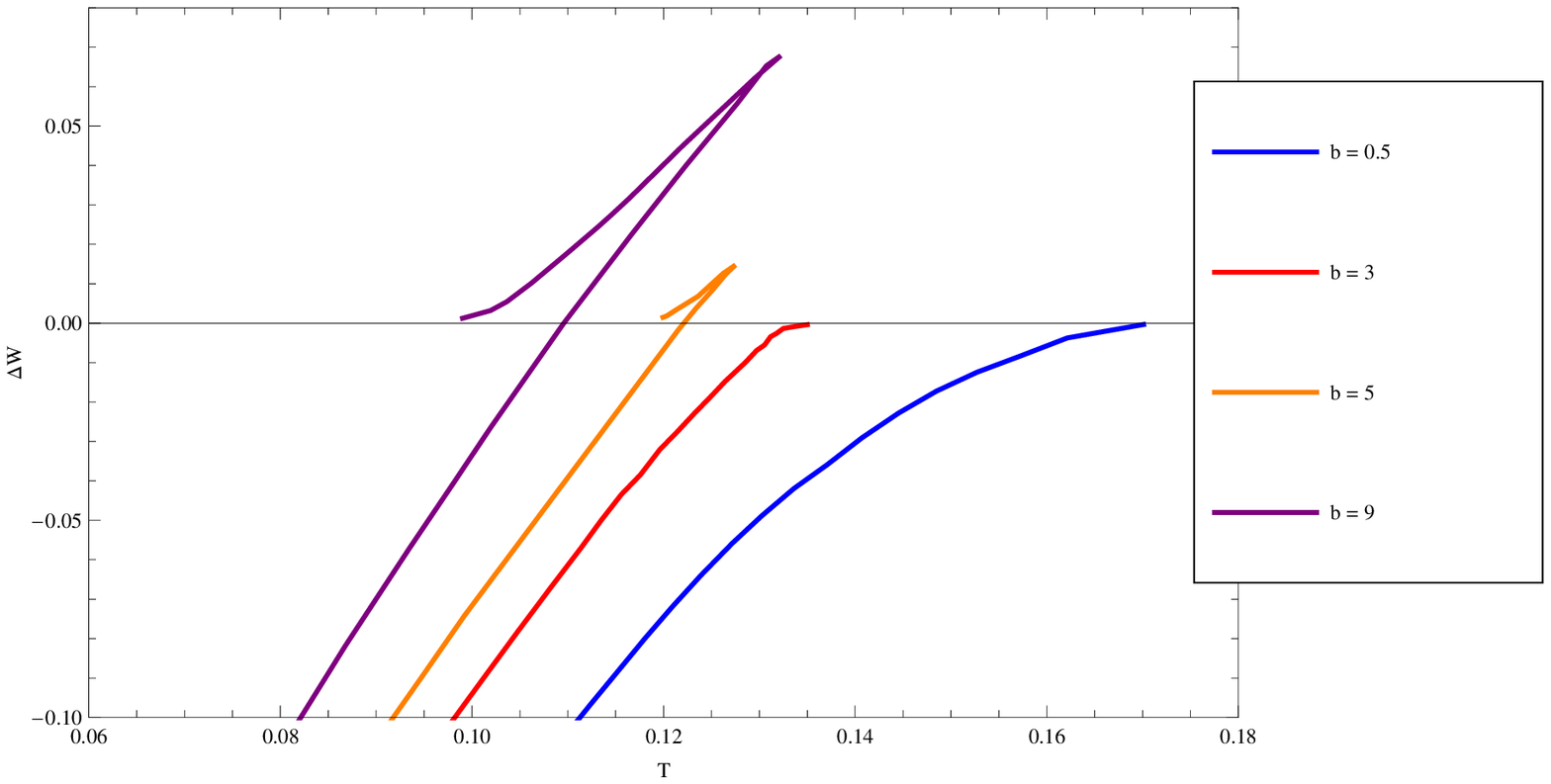}}
\caption{(a) Value of the condensate as a function of the temperature for the model with $G=(1+b\eta^2)^{-1} $, $U=1+{\eta^2\over 6},\ J=9\eta^2$, for 
 $b=0.5,\ 3,\ 5,\ 9$  (from bottom to top, the $b=0.5$ case being described by the undermost curve at $T=0$).
(b) Normalized free energy for the same model and same values of $b$ (the $b= 0.5$ case being described by the rightmost curve). 
The lower branches of the $b=5$ and $b=9$ curves correspond to the upper branches in fig. 1a.
\label{figspec1}}
\end{figure}

Figure 1a  shows the condensate $\langle O_2\rangle$ as function of the temperature for $qL=3$ and $b=0.5,\ 3,\ 5,\ 9$.
One can see that for $b$ above a critical value $b_{\rm cr}\sim 3$ there is a new branch of the solution.
Figure 2b shows the difference between the free energy of the solution and the free energy of the RNAdS black hole.
{}From this figure one can observe a number of important features.

For small values of $b$, the curves reproduce, as expected, the same qualitative features 
of the HHH model:  there is only one hairy black hole solution, and the normalized free energy (i.e. the free energy of  the  hairy solution minus the free energy of the RNAdS solution) becomes negative exactly at $T<T_c$, showing that the  black hole with scalar hair
dominates the thermodynamics in this regime.
The first derivative of the free energy is continuous at the transition, showing that the transition is second order.
The condensate approaches zero as $\langle O_2\rangle \sim (T_c-T)^\alpha$, with $\alpha \approx 1/2$ for all $b<b_{\rm cr}$.

For $b$ greater than $b_{\rm cr}$, the curves become multivalued, corresponding to the presence of two
black hole solutions with different values of $\langle O_2\rangle $ at the same temperature. 
Figure 2b  shows that the upper branch (i.e. with higher $\langle O_2\rangle $) has less free energy than the lower branch, and hence dominates the thermodynamics.

The critical temperature $T_c$ is the temperature at which the normalized free energy 
changes sign. Above $T_c$, the normalized free energy becomes positive and the RNAdS solution becomes thermodynamically favorable.
As can be seen from figs. 1a, 1b, this temperature is below
the temperature at which the lower and upper branches of the solution join. This implies that for $T>T_c$ the thermodynamically relevant solution is RNAdS,
and the vacuum expectation value has a discontinuous jump from a non-zero positive value to $\langle O_2\rangle =0$.

For $b>b_{\rm cr}$ the normalized free energy  is continuous but not differentiable at $T=T_c$, proving that the transition becomes first order.

\subsection{Model with $G(\eta )= (1+b \eta^\alpha )^{-1}$ }

This example illustrates a model with $\eta >0$ and represents another check of conjecture 4 (the first check is the model (\ref{tuna}) of \cite{Franco:2009if}
with $\beta =(\alpha-2)^{-1}$). 
We have used the value $b=0.5$, with three different values of $\alpha$, $\alpha = 3.75, \ 3.5, \ 3.25$. For simplicity --and to achieve greater numerical precision-- the figures have been obtained ignoring back reaction, but we have verified
that the incorporation of back reaction gives essentially the same results.
For these values of the parameters the model has second-order phase transitions. Figure 2a shows the order parameter $\langle O_2 \rangle $ in terms of the temperature $T$
for these three different values of $\alpha $. We can see that the critical temperature is independent of $\alpha $, $T_c \cong 0.11843 $.
Figure 2b  shows $\log \langle O_2 \rangle $ vs. $\log (1-T/T_c)$. Near the critical temperature we have the behavior
\be
\langle O_2 \rangle \sim (T_c-T)^\beta \ ,
\ee 
with $\beta \cong 0.58,\ 0.67,\ 0.81$ for $\alpha=3.75 ,\ 3.5,\ 3.25$ respectively. Remarkably, $\beta $ is related to $\alpha $ by the same law as in the model (\ref{tuna}):
\be
\beta \cong (\alpha-2)^{-1}\ .
\ee
Despite this coincidence, the two models are inequivalent (though they seem to belong to the same universality class).
 As a check of conjecture 3, we have also studied the case $\alpha =3$, which gives $\beta \cong 1$.

\begin{figure}[tbh]
\centering
\subfigure[]{\includegraphics[width=7.5cm]{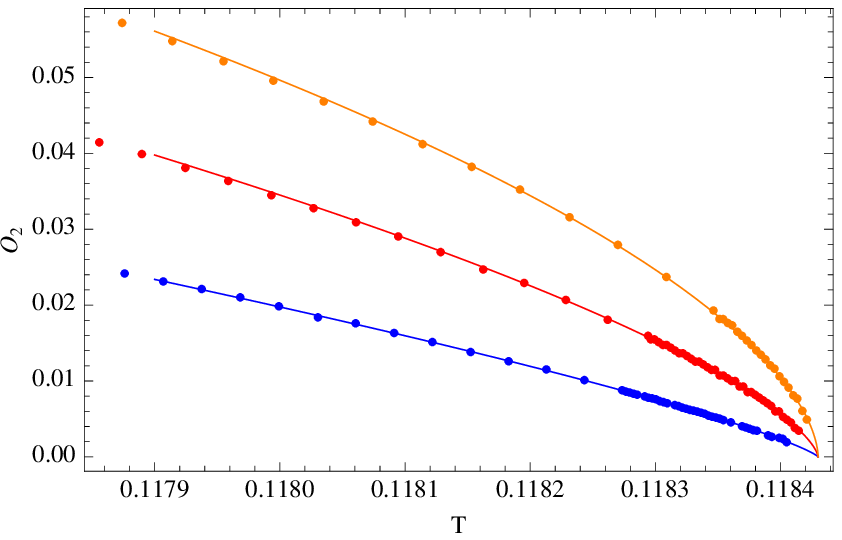}}
\subfigure[]{\includegraphics[width=7.5cm]{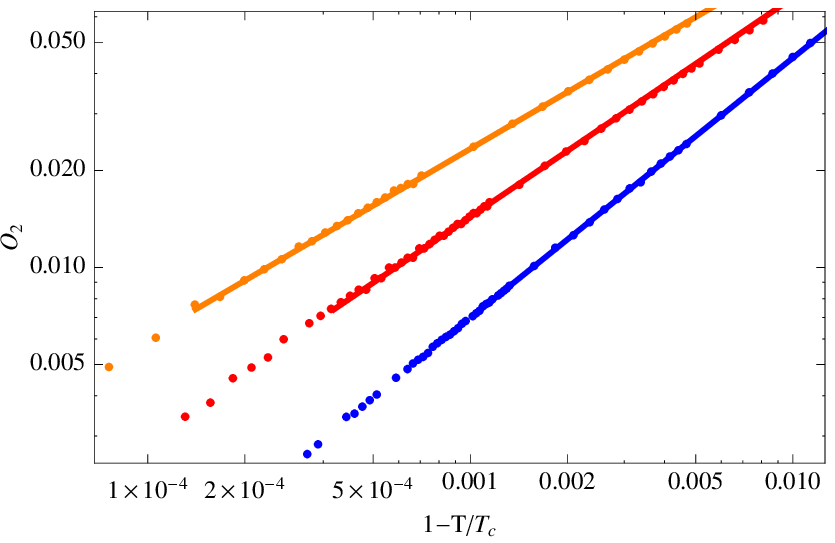}}
\caption{ (a) Value of the condensate as a function of the temperature for the model with $G= (1+0.5 \eta^\alpha )^{-1}$, $U=1+{\eta^2\over 6},\ J=\eta^2$, for  (from top to bottom)
$\alpha = 3.75 $ (orange),  $\alpha = 3.5 $ (red), $\alpha = 3.25 $  (blue).
(b) $\langle O_2 \rangle $ vs. $(1-T/T_c)$ in logarithmic scale, showing that different values of $\alpha $ have different slopes (conventions as in (a)).
\label{figspec2}}
\end{figure}

\subsection{Model with $G(\eta )=\cosh \eta $ }

Our third example illustrates the fact that there are as well non-trivial choices of $G(\eta )$ defining  models
which share essentially the same physics as the HHH model, even when the coupling becomes exponentially large at large $\eta $.
We  consider the model defined by the couplings
\be
G=\cosh \eta  \ ,\qquad U=1+{\eta^2\over 6}\ ,\qquad  J=q^2\eta^2\ .
\ee
In a sense, this model is complementary to the model of section 3.1, where $G$ was suppressed at large $\eta $.
{} If $\eta $ is small in the whole space $r_+<r<\infty $, then $\cosh \eta\sim 1$ and the couplings are  essentially the same as
in the HHH model. If $\eta_+$ is large, then $G$ becomes exponentially large near the horizon region and in principle one may expect
some different physics. The results we have found, summarized by figures 3a and 3b, are as follows.


The uncondensed $\eta =0$ phase is as usual described by the RNAdS solution.
Below some critical temperature $T_c\approx 0.188$ we  again find
a hairy black hole solution. This critical temperature is slightly above the one of the HHH model. In general, we have checked that for a model with $G(\eta )=\cosh \alpha \eta $ 
the critical temperature depends on the coefficient $\alpha $. This is consistent with the above conjecture 2.

Figure 3a  shows the value of the condensate as a function of the temperature, whereas figure 3b shows the normalized free energy.
Below the critical temperature $T_c$, a black hole solution with scalar hair appears. Figure 3b shows that this has  less free energy than the RNAdS solution, 
implying the existence of a phase transition.
The free energy has a continuous derivative at the critical point, which shows that the transition is second order.
A fit of the curve near the critical point shows that the critical exponent is  given by the 
mean field value,  $\beta \approx 1/2$, in consistency with conjecture 1.


\begin{figure}[tbh]
\centering
\subfigure[]{\includegraphics[width=7.5cm]{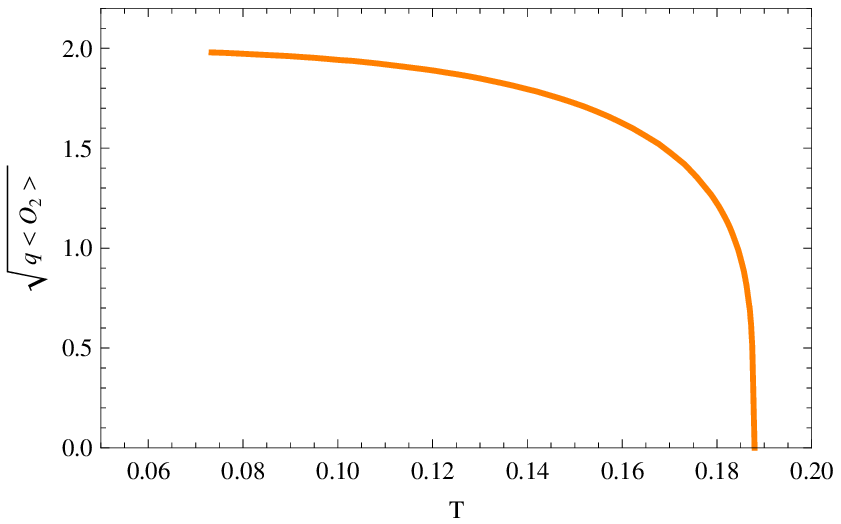}}
\subfigure[]{\includegraphics[width=7.5cm]{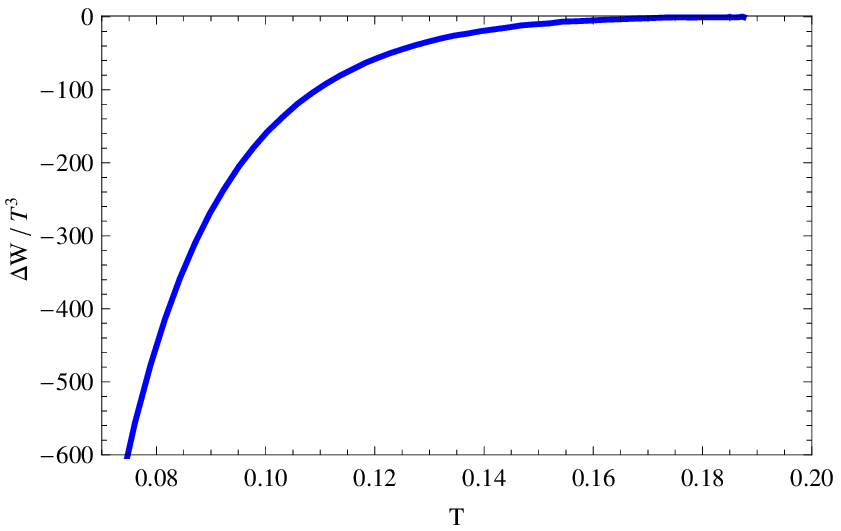}}
\caption{ (a) Value of the condensate as a function of the temperature for the model with $G=\cosh\eta $, $U=1+{\eta^2\over 6},\ J=9\eta^2$. 
(b) Normalized free energy for the same model.
\label{figspec3}}
\end{figure}

\section{Conductivity}

{} In this section we briefly discuss the set-up for studying conductivity in the general model (\ref{uno}).
We will consider $d=3$ and $G, U, J$ with the asymptotic conditions (\ref{asym}) with $m^2=-2$.
Following \cite{Gubser:2008px,Hartnoll:2008vx,Hartnoll:2008kx}, we consider time-dependent perturbations of $A_x= a_x(r)\ e^{-i\omega t}$ and $g_{tx}=f(r) \ e^{-i\omega t}$. 
In our general model, these fluctuations are governed by the following equations
\bea
&& a''_x+ \left({g'\over g}-{\chi'\over 2}+{\partial_\eta G \, \eta'\over G}\right) a_x' +
\left({\omega^2\over g^2} e^\chi - {J\over gG}\right) a_x = {\phi'\over g}\ e^\chi \Big(-f'+{2\over r}\ f\Big)\ ,
\\
&& f' -{2\over r}\ f+G\ \phi' a_x =0\ .
\eea
 Substituting the second  into the first equation, we find
 \be
a''_x+ \left({g'\over g}-{\chi'\over 2}+{\partial_\eta G \, \eta'\over G}\right) a_x' +\left( \Big({\omega^2\over g^2} -{G {\phi'}^2 \over g}\Big) e^\chi - {J\over gG}\right)a_x = 0\ .
\ee
The asymptotic behavior of the perturbations is found to be
\be
a_x = a_x^{(0)}+ {a^{(1)}_x\over r} + \ldots \ ,\qquad f=r^2 f^{(0)} + {f^{(1)}\over r}+\ldots
\ee
 The conductivity can then be found by the formula
\be
\sigma = {J_x\over E_x} = - {ia_x^{(1)}\over \omega a_x^{(0)} }\ ,
\ee
where in the second equality we have used the AdS/CFT dictionary.

At this point, we can now examine under which conditions the general class of models (\ref{uno}) lead to infinite DC (i.e. at $\omega =0$) conductivity.
The emergence of infinite DC conductivity is a general feature that follows after the $U(1)$ symmetry breaking due to a mass term for the photon.
In the present case, the Maxwell equation has a term $ {J\over gG} \ a_x$ which is the one to be interpreted as a (London) current.
As long as this is non-vanishing, one expects as usual infinite DC conductivity.
The other terms in the above Maxwell equation are due to back reaction or interaction of the Maxwell field with the geometry.
To be more precise, we need that $a_x^{(1)}/a_x^{(0)} $ is a real, non-vanishing number in the limit that $\omega\to 0$, so that
the imaginary part of the conductivity contains a pole, ${\rm Im}(\sigma ) \sim 1/\omega $. Then, by virtue of the Kramers-Kronig relation
\be
{\rm Im}(\sigma (\omega ))=-{1\over \pi}\ {\cal P}\int   { {\rm Re} (\sigma(\omega' ) )d\omega '\over \omega'-\omega }\ ,
\ee
one finds that a pole in ${\rm Im}(\sigma )$  is associated with a Dirac delta function in ${\rm Re} (\sigma(\omega ) )$.
Typically, one expects that any  $J$ which vanishes at $\eta=0$ (i.e. in the normal, AdS Reissner-Nordstr\" om phase)
and does not vanish for $\eta\neq 0$ should lead to superconductivity, if it also leads to 
a phase transition like those described in the previous sections.


\section{Concluding remarks}

Summarizing, we have presented a large class of models which can be used as a phenomenological set-up for superconductivity or for model building.
The general models interpolate between  different models appeared in the literature.
We have discussed general conditions that the couplings must satisfy for the theory to holographically describe superconducting phase transitions.
In section 3 we have also conjectured how some aspects of the dynamics of the phase transition are encoded in the couplings $G,\ U,\ J$.
The concrete examples investigated here partly confirm this picture, but a more thorough investigation 
would clearly be necessary to substantiate these conjectures, possibly by analytic, rather than numerical, techniques.
We have explored models with non-trivial $G(\eta )$. From a more fundamental standpoint, such models might  arise from M-theory or string theory compactifications,
or by quantum effects, since such couplings are compatible with $U(1)$ local symmetry and general covariance.

There are, of course,  other aspects of superconducting phase transitions which have not been
discussed here.
There may be many surprises in the space of models parametrized by the couplings $G(\eta),\ U(\eta),\ J(\eta )$ and our analysis  only explores
a small corner of this space. In particular, one feature that emerges in certain regions of the parameter space for some models is the appearance of critical curves where
the sign of the second derivative $d^2\langle O_2\rangle/dT^2$ changes at some $T<T_c$. In some cases this (probably unphysical) feature is cured by the incorporation of back-reaction,
but we have found cases where this feature persists even after including back reaction effects.

{} Another interesting problem is to look for models with special couplings $G(\eta),\ U(\eta),\ J(\eta )$ which could
allow one to find analytical solutions.
The general system (\ref{buno})--(\ref{bcuatro}) of four coupled differential equations contains two first order equations
and two second order equations (\ref{bdos}). Using the conserved charge discussed in the appendix A, one can replace the equation for $\phi $ (\ref{btres}) by a first-order one.
A natural question is then if there is some choice of couplings where all equations are first order.
 This would render the numerical analysis simpler and perhaps lead to an analytic treatment of at least some aspects of the phase transitions.
 A simple example is presented in appendix A, where
we show that indeed there is  one model which has an extra conserved charge and the equation for $\eta $ can be substituted by a first-order equation.
 Although this particular model does not seem to be relevant for holographic superconductivity, it would be interesting to see if there are other choices
 of $G(\eta),\ U(\eta),\ J(\eta )$ where the equations can also be integrated to a first-order form (see also the superpotential method described in appendix B).

\section*{Acknowledgements}

We would like to thank S. Franco and D. Rodriguez-G\'omez for useful discussions.
F.A. is supported by a MEC FPU Grant No.AP2008-04553. He would also like to thank 
the Galileo Galilei Institute for Theoretical Physics for the hospitality and the INFN for partial support during the completion of this work.
We acknowledge support by MCYT  Research Grant No.  FPA 2007-66665 and Generalitat de
Catalunya under project 2009SGR502.

\renewcommand{\thesection}{\Alph{section}}
\setcounter{section}{0}

\section{General ansatz and conserved charges}

Consider the following ansatz
\be
ds^2= -g(r) e^{-\chi (r)} dt^2 +{dr^2\over g(r)} + a^2(r)\ (dx^2+dy^2)\ ,\qquad A=\phi(r)dt\ ,\qquad \eta=\eta(r)\ .
\ee
which is slightly more general than the ansatz (\ref{arass}) adopted in section 2.2.
The Lagrangian (\ref{uno}) takes the form
\be\label{L2}
\sqrt{-g} {\cal L} = 2 e^{-{\chi\over 2}} a' \Big( (a g)'-ag \chi' \Big)+ {a^2\over 2} G(\eta ) e^{\chi\over 2} {\phi'}^2+{6a^2\over L^2}  e^{-{\chi\over 2}} U(\eta ) -{a^2\over 2} e^{-{\chi\over 2}} g {\eta'}^2
+{a^2\over 2g} e^{\chi\over  2} J(\eta ) \phi^2
\ee
The equations of motion are given by
\bea
&& 2a''+a'\chi' + {a\over 2} {\eta'}^2+ {a\over 2g^2} e^\chi J(\eta)\phi^2=0\ ,
\label{auno}
\\
&& {1\over 4}\ {\eta'}^2+ {G(\eta)\over 4g}\ e^\chi {\phi'}^2+{a'g'\over ag} +{{a'}^2\over a^2}+{2a''\over a } -{3\over L^2 g}\ U(\eta) +{1\over 4g^2}\ e^\chi J(\eta) \phi^2=0\ ,
\label{ados}
\\
&& \phi''+\phi' \ \left( {2a'\over a} +{\chi'\over 2} +{\p_\eta G \eta'\over G} \right)- {J(\eta)\over g G(\eta)} \ \phi =0\ ,
\label{atres}
\\
&& \eta'' +\eta'  \ \left( {2a'\over a} -{\chi'\over 2} +{g'\over g} \right)+ {1\over 2g}\ e^\chi \p_\eta G \ {\phi'}^2 + {6\over L^2g}\ \p_\eta U + {1\over 2g^2}\ e^\chi \p_\eta J\ {\phi}^2 =0\ .
\label{acuatro}
\eea

One can show, generalizing the similar derivation given in \cite{Gubser2}, that the following quantity 
\be
Q_1= a^2 e^{\chi /2}  \left( a^2 \left( {e^{-\chi} g\over a^2}\right)' - G(\eta )\phi\phi'\right)
\label{Qcon}
\ee
is a constant. 
This is a conserved charge associated with the following scaling symmetry
\be
\phi\to c^2\phi \  ,\qquad a\to {a\over c}\ ,\qquad \chi\to \chi - 4\log(c)\ ,
\ee
\be
g\to g\\ ,\qquad \eta\to\eta\ .
\ee
A natural question is if there is a particular choice of the couplings $G,U,J$ for which
the system has some extra conserved charge.
We found that the system with 
\be
G=e^{\alpha\eta}\ , \qquad J=q^2 e^{\alpha\eta}\ , \qquad  U=1\ ,
\label{expo}
\ee
enjoys the following scaling symmetry:
\be
\phi\rightarrow c\phi\ ,\qquad \quad \eta\rightarrow \eta -{2\over \alpha} \log c\ ,
\ee
with trivial transformation rules for $\chi,\ a,\ g$, under which $G\rightarrow c^{-2}G ,\  J\rightarrow c^{-2}J $ and
the action remains invariant. 
The corresponding Noether charge is given by
\be
Q_2= a^2 e^{\chi /2}G(\eta)\phi\phi'+ {2\over \alpha }\ a^2e^{-{\chi\over 2}}g\eta'    \ .  
\label{Qccon}
\ee
It is easy to check that eqs. (\ref{Qcon}) and (\ref{Qccon}) are integrals  of the equations of motion.
Combining (\ref{Qcon}) and (\ref{Qccon}) one finds that, in the model (\ref{expo}),  the following combination is also conserved
\be
Q_3\equiv Q_1+Q_2= a^4 e^{\chi /2}   \left( {e^{-\chi} g\over a^2}\right)' + {2\over \alpha }\ a^2e^{-{\chi\over 2}}g\eta'\ .
\label{Qzon}
\ee
By choosing the gauge $a=r$ the equations (\ref{auno}) and (\ref{ados}) become first order. Thus, in this gauge, one is left with a system of four coupled, independent, first order differential equations,
(\ref{auno}) , (\ref{ados}), (\ref{Qcon}) and  (\ref{Qzon}).
 
One could try to use this first-order system to find, if not  analytic, at least numerically simpler solutions for $\eta $ and $\phi $.
Unfortunately, this model does not seem to exhibit holographic superconductivity. The basic reason is that any solution  asymptotic to AdS 
with $\eta\to \eta_0$ (in particular, $\eta\to 0$) at infinity gives a mass to the photon; see eq. (\ref{uno}).

\section{AdS Reissner-Nordstr\" om solution via superpotential}\label{appA}

Here we provide a simple derivation of the AdS Reissner-Nordstr\" om  solution using the superpotential method. 
We hope that this type of approach may also be useful for the search of analytic  black hole solutions with non-trivial scalar hair $\eta $.
In general, one considers  a Lagrangian of the type
\be
\mathcal{L}=\frac{1}{2}G_{ij}\frac{d\a^i}{dr}\frac{d\a^j}{dr}-V(\a^i)\ ,
\ee 
where $G_{ij}$ is a symmetric matrix, that may depend on the fields $\a^i$. The equations of motion are given by
\be
\frac{d}{dr}\Big[G_{ij}\frac{d\a^j}{dr}\Big]=\frac{1}{2}\Big(\frac{\p}{\p\a^i}G_{jk}\Big)\frac{d\a^j}{dr}\frac{d\a^k}{dr}-\frac{\p V}{\p\a^i} \ .
\ee
One then assumes that a superpotential $W(\a^i)$ exists, which is such that it satisfies the following relation:
\be\label{superp1}
V=-\frac{1}{2}G^{ij}\frac{\p W}{\p\a^i}\frac{\p W}{\p\a^j} \ .
\ee
If $W$ is found, then one can show that the first order system,
\be\label{superp2}
\frac{d\a^i}{dr}=\mp G^{ij}\frac{\p W}{\p\a^j}
\ee
automatically solves the equations of motion (the converse is in general not true; not all solutions of the second-order system solve
the first-order system).

We now apply this procedure to our case.
We consider the ansatz
\be
ds^2=-{S\over R} dt^2+ {R\over S} dr^2+ R^2 (dx^2+dy^2)\ ,\qquad A_{\mu}dx^{\mu}=\phi(r)\ dt \ ,
\ee
where $S=S(r),\ R=R(r)$ and $\eta =\eta (r)$. The Lagrangian then becomes
\be
\sqrt{-g}{\cal L} =2 R' S'+ {R^2\over 2}\ G(\eta ) {\phi' }^2+  {6\over L^2} R^2\ U(\eta ) - {1\over 2} RS {\eta'}^2+ {1\over 2} {R^3\over S} J(\eta ) \phi^2\ .
\ee
The equations of motion are given by
\bea
&&{R'S'\over RS} =- {1\over 4} {R\over S} G \phi'^2+ {1\over 4} {\eta'}^2+{3\over L^2} {R\over S}\ U+ {1\over 4}\ {R^2\over S^2}\ J\phi^2\ ,
\label{cuno}\\
&& R''+{R\over 4} {\eta'}^2+ {R^3\over 4 S^2} J(\eta)\phi^2 =0\ ,
\label{cdos}\\
&& \phi''+\phi' \left( {2R'\over R}+ {\p_\eta G \ \eta'\over G}\right) -{R J\over SG}\ \phi =0\ ,
\label{ctres}\\
&&\eta''+\Big( {R'\over R} +{S'\over S} \Big)\ \eta' +{R\over 2S}\ \p_\eta G\ {\phi'}^2 + {6\over L^2}\ {R\over S} \ \p_\eta U+ {R^2\over 2S^2}\p_\eta J \ \phi^2=0 \ .
\label{ccuatro}
\eea
Combining eqs. (\ref{buno}) and (\ref{bdos}) one can write the simpler relation
\be
{1\over R^2} \ \left( R'S \right)' = {3 \over L^2}\  U(\eta) -   { 1\over 4}  \ G(\eta){\phi'}^2\ .
\ee


The relation (\ref{superp1}) becomes 
\be
\p_R W\p_S W+\frac{1}{R^2 G(\eta )}(\p_\phi W)^2  - \frac{1}{RS }(\p_\eta W)^2    =12 R^2 U(\eta ) +\frac{R^3}{S}\, J(\eta )\phi^2\ .
\label{WRSh}
\ee
A superpotential satisfying this relation can be found for $\eta =0$.
We choose units such that $L=1$.
In this case (\ref{WRSh}) reduces to (we assume $U(0)=G(0)=1$)
\be
\p_R W\p_S W+\frac{1}{R^2 }(\p_\phi W)^2=12 R^2\ .
\label{WRS}
\ee
We have found two simple solutions to eq. (\ref{WRS})
\bea
W_1=2(R^3+S)-\frac{R}{2}(\phi-\mu)^2\ ,
\\
W_2= 4 \sqrt{ R^3-M } \sqrt{S-{1\over 4 } \ R (\phi -\mu )^2 }\ .
\eea
It is easy to see that the AdS Reissner-Nordstr\" om  solution arises from these superpotentials. Consider, for example, the case of $W_1$. 
The first order system (\ref{superp2}) then becomes
\be
R' =1\ ,\qquad S' =3 R^2-{1\over 4}(\phi-\mu)^2\ ,\qquad \phi' = - {1\over R} \ (\phi -\mu ) \ .
\ee 
This is easily solved giving
\be
R= r\ ,\qquad S=r^3-M+{\rho^2\over 4 r}\ ,\qquad \phi=\mu-{\rho\over r}\ ,
\ee
where $M$ and $\rho$ are an integration constants. This reproduces the RNAdS solution (\ref{RNg}).
Clearly, it would be  interesting to find a generalization of $W_1$ or $W_2$ to include $\eta $ dependence.
This could allow one to have more analytic control over the hairy black hole solution and thus over the properties of the
phase transition.

\end{document}